\documentclass[conference]{IEEEtran}
\IEEEoverridecommandlockouts
\usepackage{cite}
\usepackage{amsmath,amssymb,amsfonts}
\usepackage{algorithmic}
\usepackage{graphicx}
\usepackage{textcomp}
\usepackage{xcolor} 

\usepackage{siunitx} 
\usepackage{bm} 
\usepackage{circuitikz}
\usepackage{booktabs} 
\DeclareSIUnit \voltampere { VA }
\DeclareSIUnit \var { var }
\DeclareSIUnit \pu {p.u.}
\usepackage{subfigure}
\usepackage{float}

\newcommand{\ma}[1]{\left[\begin{matrix} #1 \end{matrix}\right]} 
\newcommand{\Eq}[1]{Eq. (\ref{#1})}

\newcommand{\ie}{\emph{i.e.},~}

\newcommand{\pt}{\Delta}

\def\BibTeX{{\rm B\kern-.05em{\sc i\kern-.025em b}\kern-.08em
    T\kern-.1667em\lower.7ex\hbox{E}\kern-.125emX}}
\begin{document}

\title{Robust Grid-Forming Control Based on Virtual Flux Observer
}

\author{\IEEEauthorblockN{Xueqing Gao}
\IEEEauthorblockA{\textit{Global College} \\
\textit{Shanghai Jiao Tong University}\\
Shanghai, China \\
gaoxq113@sjtu.edu.cn}
\and
\IEEEauthorblockN{Jun Zhang}
\IEEEauthorblockA{\textit{Global College} \\
\textit{Shanghai Jiao Tong University}\\
Shanghai, China \\
zhangjun12@sjtu.edu.cn}
\and
\IEEEauthorblockN{Tao Li}
\IEEEauthorblockA{\textit{Spintrol} \\
Shanghai, China \\
tao.li@spintrol.com}
\and
\IEEEauthorblockN{Mingming Zhang}
\IEEEauthorblockA{\textit{Student Innovation Center} \\
\textit{Shanghai Jiao Tong University}\\
Shanghai, China \\
brucechang@sjtu.edu.cn}
}

\maketitle

\begin{abstract}
This paper investigates the design and analysis of a novel grid-forming (GFM) control method for grid-connected converters (GCCs). The core novelty lies in a virtual flux observer-based synchronization and load angle control method. The terminal voltage of the converter is directly regulated to provide voltage-source behavior. The control parameters are designed for decoupling and pole placement. The proposed method exhibits strong robustness in stability and dynamical performance across varying and uncertain grid strengths. The robust control performance of the proposed method is first demonstrated by small-signal analysis, then validated by experiments on a \SI{20}{\kilo\voltampere} power conversion system. 
\end{abstract}

\begin{IEEEkeywords}
grid-forming control, grid synchronization, virtual flux observer, stability, robustness, sensorless  
\end{IEEEkeywords}

\section{Introduction}
In modern power systems with a growing share of distributed generation (DG), grid connected converters (GCCs) controlled by traditional grid-following (GFL) strategies face increasing stability risks, particularly under weak grid conditions or parallel operation~\cite{ORathnayake2021}. The grid-forming (GFM) controllers have therefore been suggested as an effective solution for stable integration of DG units at weak points of common coupling (PCC)~\cite{ORosso2021,OLasseter2020}. 

Two defining characteristics of GFM converters are their voltage-source behavior and strong coupling between active power and grid synchronization. The former is realized by direct manipulation of converter terminal voltage, while the latter is rooted in power-synchronization control (PSC), which emulates the droop mechanism of synchronous generators~\cite{Zhang2010,Robust2019}. Although voltage-source behavior and PSC enables stable operation under weak grid condition, recent studies have revealed that GFM controllers also lack stability robustness~\cite{Wang2020,Li2022}. The most notable issue is the degradation of control performance under strong grid condition, including limited bandwidth and low frequency oscillation, which are resulted by increased sensitivity of active power to load angle perturbations~\cite{Zhao2024l,Gao2024}. 

Various remedies have been proposed to enhance stability robustness of GFM controllers with respect to grid strength. For example, current reference feedforward~\cite{RFPSC2020} or implicit current loop~\cite{Zhao2024} are incorporated into PCC voltage controller to improve dynamic performance under strong grid condition; in~\cite{Universal2021}, PSC is hybridized with phase-locked loop (PLL), which can be used for synchronization under strong grid condition to ensure stable operation. However, these approaches merely refine existing control loops without changing the nature of power synchronization, and may further complicate parameter tuning and dynamic performance of closed-loop system~\cite{LiuUVC2024,Zhao2024l}. 

Recently, some disturbance observer-based GFM controllers have been proposed, which obtained satisfying performances in sensorless and multifunctional GCC control~\cite{Nurminen2023,Nurminen2024,Nurminen2025}. However, these methods still adopt the traditional PSC for synchronization, or adopt arbitrarily defined reference frames. Therefore, the stability robustness limitation of existing GFM controller is not thoroughly addressed. 

Potentially, grid synchronization can be incorporated in observers based on mathematical model of the GCC. In contrast to non-model-based synchronization mechanisms of GFL and GFM, observer-based synchronization makes full use of available model information, such as filter topology and nominal impedance. Therefore, stability margin and dynamic
performance can be actively shaped, which potentially
overcomes the stability robustness limitations of existing GCC control paradigms. 

Conceptually, the task of grid synchronization is much similar to position estimation in electrical drives. Moreover, the mathematical models of permanent-magnet synchronous machines (PMSMs) and grid-connected converters (GCCs) share strong structural similarity, which facilitates the transfer of mature PMSM sensorless control techniques~\cite{Marko2018,Flux2022}, such as flux observer, into GCC applications. Therefore, this article makes an attempt at improving the stability robustness of GFM controllers with virtual flux observer-based synchronization.



\section{Mathematical modeling of the GCC}

\begin{figure}[!tbp] 
\centering 
\includegraphics[width=0.5\textwidth,trim={8cm 3.5cm 3cm 3.5cm},clip]{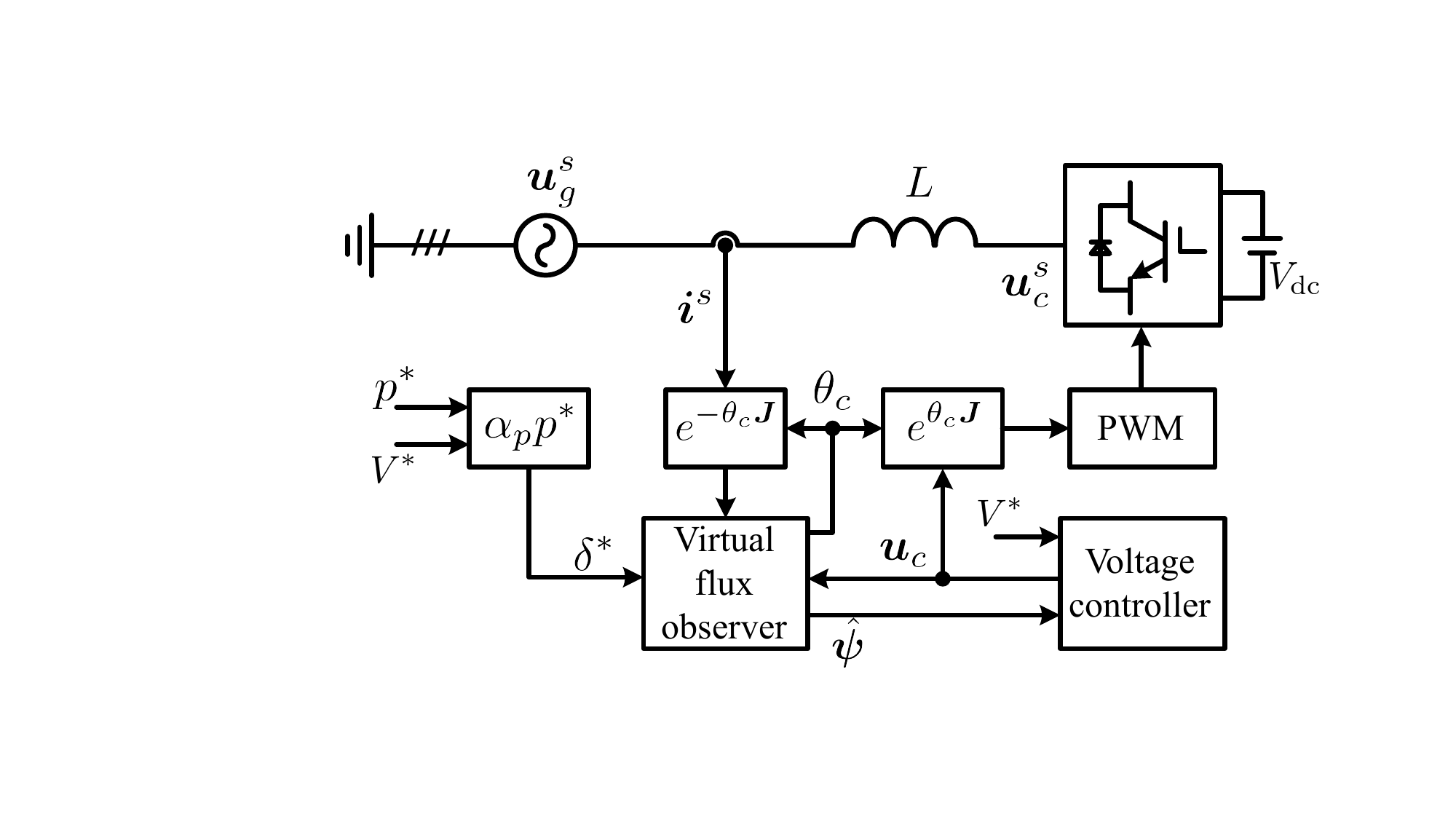}
\caption{Circuit topology of GCC and block diagram of the proposed GFM controller. Superscript $s$ denotes variables in $\alpha\beta$-frame. }\label{fig:fig1}
\end{figure}

Fig.~\ref{fig:fig1} shows the circuit topology and block diagram of the proposed GFM converter, where $L$ is the total impedance including filter impedance and grid impedance. 

\subsection{Current model of the GCC}
The mathematical model of the GCC in two-phase stationary frame ($\alpha\beta$-frame) is 
\begin{equation}\label{eq:iab}
\begin{aligned}
    \dot{\bm i}^s&=(\bm u_c^s-\bm u_g^s)/L, \;\bm u_g^s=e^{\theta_g\bm J}\bm u_g^g, \\
    \dot{\theta}_g&=\omega_g,\;\bm u_g^g=\ma{U_g&0}^T, 
    \end{aligned} 
\end{equation}
where $\bm i^s$ is converter output current, $\bm u_c^s$ is converter output voltage, $\bm u_g^s$ is grid voltage with frequency $\omega_g$, phase angle $\theta_g$ and amplitude $U_g$. 

The variables are then transformed into the controller coordinate frame ($dq$-frame), which rotates at frequency $\omega_c$ and phase angle $\theta_c$ as shown in Fig.~\ref{fig:fig2}, 
\begin{equation}\label{eq:cvariables}
    \bm i=e^{-\theta_c\bm J}\bm i^s,\;\bm u_c=e^{-\theta_c\bm J}\bm u_c^s,\;\bm u_g=e^{-\theta_c\bm J}\bm u_g^s. 
\end{equation}
Define the angle difference between the grid and controller reference frames as $\delta=\theta_c-\theta_g$, and substitute~\Eq{eq:cvariables} into~\Eq{eq:iab}, the mathematical model of GCC in $dq$-frame is 
\begin{equation}\label{eq:ic}
\begin{aligned} 
    \dot{\bm i}&=-\omega_c\bm J\bm i+(\bm u_c-\bm u_g)/L, \\
    \dot{\delta}&=\omega_c-\omega_g, \;
    \bm u_g=e^{-\delta\bm J}\bm u_g^g, 
    \end{aligned} 
\end{equation}
where $\delta$ is included as a state variable. 

Active power and PCC voltage magnitude are defined as 
\begin{subequations}
\begin{align}
    p&=\kappa\bm u_g^T\bm i,\;\kappa=3/2, \label{eq:p}\\
    V&=(\bm u_c^T\bm u_c)^{1/2}. \label{eq:V}
    \end{align}
\end{subequations}

\subsection{Virtual flux model of the GCC}
The virtual flux of GCC is defined as 
\begin{equation}\label{eq:psi}
    \bm \psi=L\bm i+\bm \psi_g, \;\bm \psi_g=(\omega_g\bm J)^{-1}\bm u_g. 
\end{equation}
Substitute~\Eq{eq:psi} into~\Eq{eq:ic}, the dynamics of virtual flux is 
\begin{equation}\label{eq:psidot}
    \dot{\bm \psi}=-\omega_c\bm J\bm \psi+\bm u_c. 
\end{equation} 


\section{Synchronization through virtual flux observer}
From Section II. B, the position of grid voltage vector is contained in virtual flux vector. Therefore, in this section, a synchronous rotating frame is obtained from virtual flux observer for the purpose of active power control. 
\subsection{Operating points in controller coordinate frame} 

\begin{figure}[!tbp] 
\centering 
\includegraphics[width=0.5\textwidth,trim={5.4cm 4.5cm 4cm 4cm},clip]{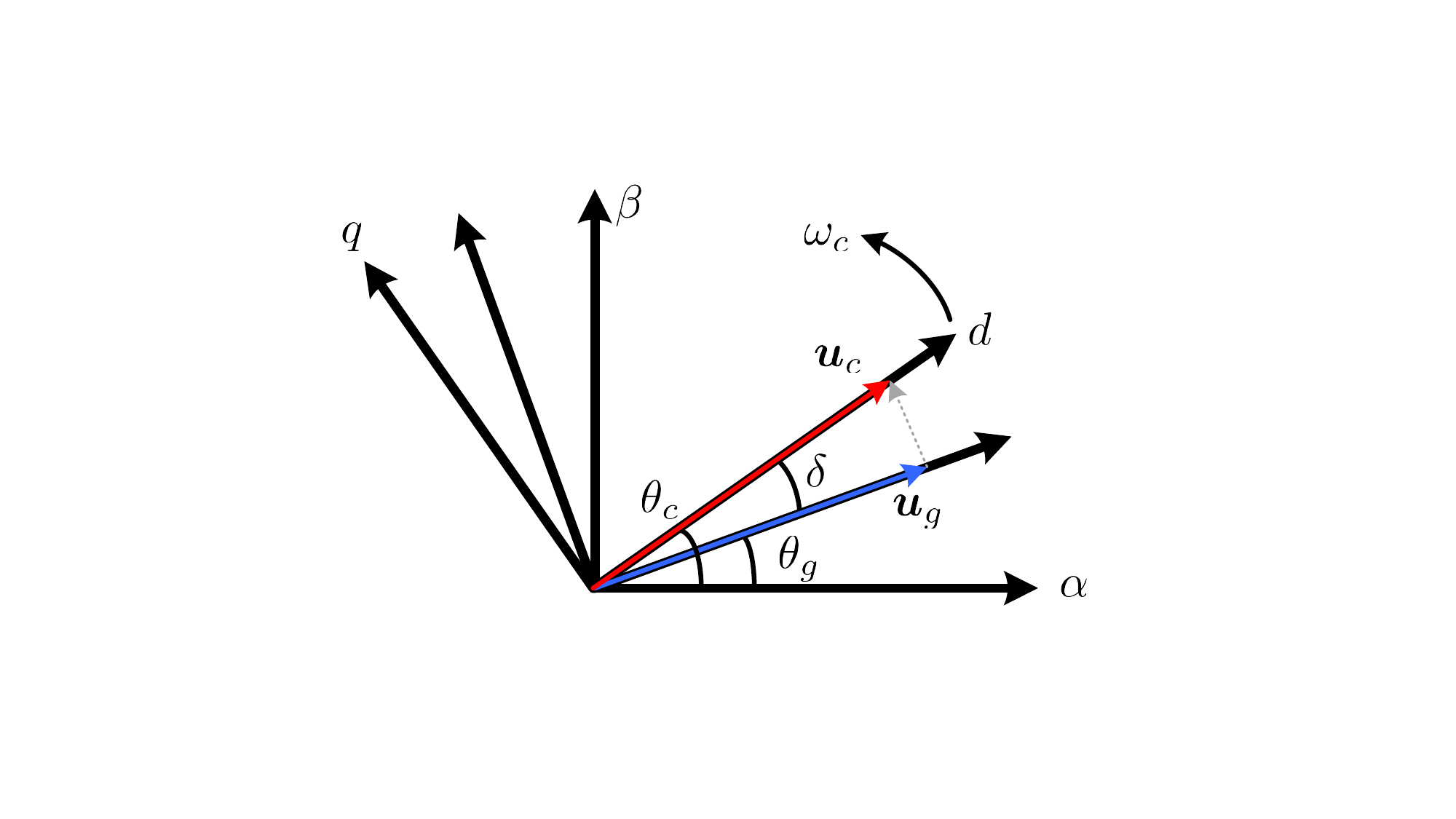}
\caption{Definition of controller coordinate frame and load angle $\delta$. }\label{fig:fig2}
\end{figure} 

First, the desired operating points of $\bm u_c$, $\bm u_g$ and controller frame position $\delta$ in Fig.~\ref{fig:fig2} corresponding to active power set-point $p^*$ and voltage amplitude set-point $V^*$ are specified. 

For simplicity, the steady-state of converter voltage aligns with $d$-axis in steady-state, \ie $\bm u_c^*=[V^*;0]$. To maintain synchronization with the grid voltage, it is required that $\omega_c^*=\omega_g$. Then, let the right hand side of~\Eq{eq:psidot} equals $\bm 0$, the desired steady-state of virtual flux is $\bm \psi^*=(\omega_g\bm J)^{-1}\bm u_c^*$. 

Let the load angle corresponding to $p^*$ be $\delta^*$, then $\bm u_g^*=e^{-\delta^*\bm J}\bm u_g^g$. Substitute~\Eq{eq:psi} into~\Eq{eq:p} with steady-state values, 
\begin{equation}
    p^*=\kappa\bm u_g^{*T}\bm \psi^*/L=\frac{\kappa U_gV^*}{\omega_{g}L}\sin{\delta^*}. 
\end{equation} 
Therefore, $\delta^*$ is almost linearly related to $p^*$, 
\begin{equation}\label{eq:ptodelta}
    \delta^*=\sin^{-1}{(\alpha_pp^*)}\approx \alpha_pp^*, \;\alpha_p=\frac{\omega_{g}L}{\kappa U_gV^*}. 
\end{equation}

\subsection{Virtual flux observer} 
The virtual flux observer is constructed from~\Eq{eq:psidot} as 
\begin{subequations}\label{eq:ob}
\begin{align} 
\dot{\hat{\bm \psi}}&=-\omega_c\bm J\hat{\bm \psi}+\bm u_c+\bm K_o\bm e, \label{eq:ob1}\\
\bm e&=L\bm i+\bm \psi_g^*-\hat{\bm \psi}, \;\bm \psi_g^*=(\omega_0\bm J)^{-1}\bm u_g^*, \label{eq:ob2}
\end{align} 
\end{subequations}
where $\omega_0$ is the nominal grid frequency.

The grid frequency estimation $\omega_c$ is given by a proportional-integral (PI) regulator driven by $\bm e$, 
\begin{equation}\label{eq:wc}
\begin{aligned} 
\dot{\bm \gamma}&=\bm e, \;
\omega_c=\bm k_i\bm \gamma+\bm k_p\bm e. 
\end{aligned}
\end{equation}
The phase angle of controller coordinate frame is $\dot{\theta}_c=\omega_c$. 

\subsection{Error dynamics and gain design}
The estimation errors are defined as 
\begin{equation}
    \tilde{\bm \psi}=\bm \psi-\hat{\bm \psi},\;\tilde{\theta}=\delta-\delta^*,\;\tilde{\omega}=\omega_c-\omega_g. 
\end{equation}
Their dynamic equations can be derived as 
\begin{subequations}\label{eq:oberror}
    \begin{align}
        \dot{\tilde{\bm \psi}}&=-(\omega_g+\tilde{\omega})\bm J\tilde{\bm \psi}-\bm K_o\bm e, \label{eq:dotpsitilde}\\
        \dot{\tilde{\theta}}&=\tilde{\omega}, \\
        \dot{\tilde{\omega}}&=\bm k_i\bm e+\bm k_p\dot{\bm e}\label{eq:oberror3}, 
    \end{align}
\end{subequations}
where~\Eq{eq:oberror3} holds with the assumption $\dot{\omega}_g=0$.

~\Eq{eq:oberror} is a nonlinear system with state variables $\bm x=[\tilde{\bm \psi};\tilde{\theta};\tilde{\omega}]$. It can be linearized around desired equilibrium $\bm x_0^*=[\bm 0;0;0]$. First, consider $\bm e$ as an input, and obtain 
\begin{subequations}
    \begin{align}
        \pt\dot{\tilde{\bm \psi}}&=-\omega_0\bm J\pt\tilde{\bm \psi}-\bm K_o\pt\bm e,\label{eq:l1_oberror_1} \\
        \pt\dot{\tilde{\theta}}&=\pt\tilde{\omega}, \label{eq:l1_oberror_2}\\
        \pt\dot{\tilde{\omega}}&=\bm k_i\pt\bm e+\bm k_p\pt\dot{\bm e}. \label{eq:l1_oberror_3}
    \end{align}
\end{subequations} 
where $\Delta$ denotes perturbations. Then,~\Eq{eq:ob2} is linearized around $\bm x_0^*$ as 
\begin{equation}\label{eq:pte}
    \pt\bm e=\pt\tilde{\bm \psi}+\bm J\bm \psi_g^*\pt\tilde{\theta}. 
\end{equation}
Substitute~\Eq{eq:pte} into~\Eq{eq:l1_oberror_1}, the dynamic equation of virtual flux estimation error becomes 
\begin{equation}
\begin{aligned}
    \pt\dot{\tilde{\bm \psi}}&=-(\omega_0\bm J+\bm K_o)\pt\tilde{\bm \psi}-\bm K_o\bm J\bm \psi_g^*\pt\tilde{\theta}, 
    \end{aligned}
\end{equation}
where the second term can be canceled by choosing $\bm K_o=\bm k_o\bm \psi_{g}^{*T}$. This gain design decouples the dynamics of synchronization from that of virtual flux estimation, resulting in 
\begin{equation}\label{eq:ptdotpsitilde}
    \pt\dot{\tilde{\bm \psi}}=-(\omega_0\bm J+\bm K_o)\pt\tilde{\bm \psi}. 
\end{equation}
The gain vector $\bm k_o$ can be calculated by solving a pole placement problem defined as 
\begin{equation}\label{eq:sigmao}
    \text{eig}(-\omega_0\bm J-\bm k_o\bm \psi_g^{*T})=\bm \sigma_o, 
\end{equation}
where $\bm \sigma_o$ specifies two poles on left half of $s$-plane. 

Decoupling properties can also be achieved for grid frequency estimation. Substitute~\Eq{eq:pte} into~\Eq{eq:l1_oberror_3}, 
\begin{equation}\label{eq:23}
    \begin{aligned}
        \pt\dot{\tilde{\omega}}&=(\bm k_i\pt\tilde{\bm \psi}+\bm k_p\pt\dot{\tilde{\bm \psi}})+\bm k_i\bm J\bm \psi_g^*\pt\tilde{\theta}+\bm k_p\bm J\bm \psi_g^*\pt\tilde{\omega}. 
    \end{aligned}
\end{equation}
Substitute~\Eq{eq:ptdotpsitilde} in~\Eq{eq:23}, the first term of~\Eq{eq:23} can be canceled if $\bm k_p$ and $\bm k_i$ satisfies 
\begin{equation}
    \left(\bm k_i-\bm k_p(\omega_0\bm J+\bm K_o)\right)\pt\tilde{\bm \psi}\equiv\bm 0. 
\end{equation} 
Therefore, gains of the PI regulator satisfying 
\begin{equation}\label{eq:kpki}
    \bm k_i=\bm k_p(\omega_0\bm J+\bm K_o)
\end{equation} 
decouples the error convergence dynamics of virtual flux estimation from that of frequency estimation. 
\subsection{Synchronization properties}

\begin{figure}[!tbp]
\includegraphics[width=0.5\textwidth,trim={3cm 6cm 2cm 6.5cm},clip]{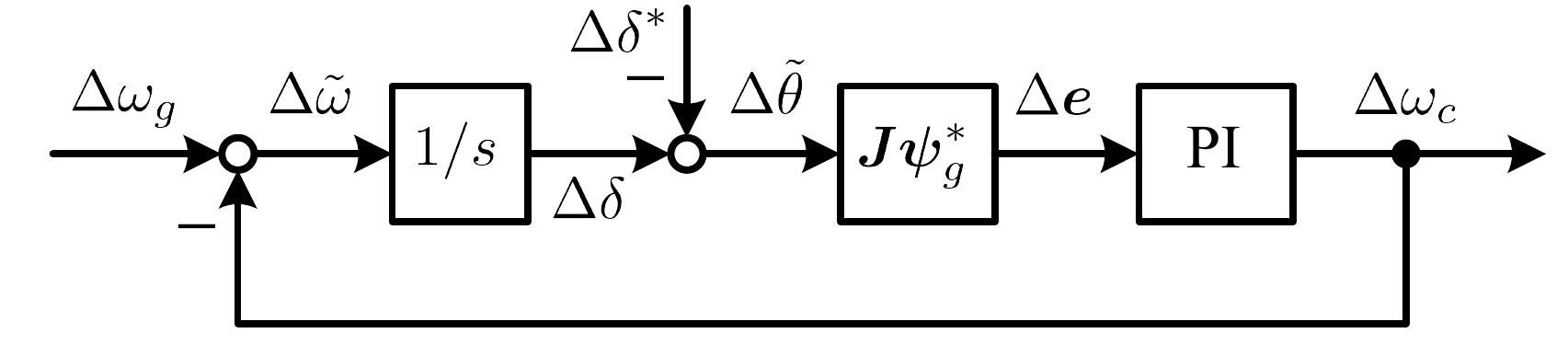}
\caption{Small-signal model of grid synchronization after decoupling. }\label{fig:3}
\end{figure}

The small-signal model of synchronization after decoupling is shown in Fig.~\ref{fig:3}, where $\pt\omega_g$ is considered as a disturbance input around nominal frequency $\omega_0$, \ie $\omega_g=\omega_0+\pt\omega_g$. Using $\pt\tilde{\theta}=\pt\delta-\pt\delta^*$ and $\pt\tilde{\omega}=\pt\omega_c-\pt\omega_g$, the dynamics of synchronization is described by the transfer function matrix 
\begin{equation}\label{eq:sync_TFM}
    \begin{aligned}
        \ma{\pt\delta(s)\\\pt\omega_c(s)}&=\frac{1}{D_1(s)}\ma{-a_2&-s\\-a_2s&-a_1s-a_2}\ma{\pt \delta^*(s)\\\pt\omega_g(s)}, \\
        a_1&=\bm k_p\bm J\bm \psi_g^*,\;a_2=-\omega_0\bm k_p\bm \psi_g^*. 
    \end{aligned}
\end{equation}
The characteristic polynomial is $D_1(s)=s^2-a_1s-a_2$. By final value theorem, the static gain of~\Eq{eq:sync_TFM} is 
\begin{equation}
    \ma{\pt\delta(\infty)\\\pt\omega_c(\infty)}=\ma{1&0\\0&1}\ma{\pt\delta^*\\\pt\omega_g}, 
\end{equation}
which ensures that the controller coordinate frame converges to the grid reference frame with the same rotational speed $\omega_g$ and desired load angle $\delta^*$. 

The dynamic performance of synchronization is adjustable by tuning the gain vector $\bm k_p$. One simple way is to specify desired bandwidth $\omega_s$ and damping ratio $\zeta$ of the second-order characteristic polynomial. Let 
\begin{equation}
    D_1(s)=s^2-a_1s-a_2=s^2+2\zeta\omega_ss+\omega_s^2, 
\end{equation} 
the corresponding gain $\bm k_p$ can be calculated as 
\begin{equation}\label{eq:kp}
    \bm k_p=\left(\ma{\psi_{g}^{*q}&-\psi_{g}^{*d}\\\psi_{g}^{*d}&\psi_{g}^{*q}}^{-1}\ma{2\zeta\omega_s\\\omega_s^2/\omega_{0}}\right)^T. 
\end{equation} 

\section{Voltage control and Closed-loop system analysis} 
\subsection{Voltage control law and gain design} 
The converter output voltage is designed as 
\begin{equation}\label{eq:uc}
    \bm u_c=\bm u_{c}^*+\bm k_v(V^*-\hat{V}), \;\hat{V}=\omega_c\left(\hat{\bm \psi}^T\hat{\bm \psi}\right)^{1/2}. 
\end{equation}

To study parameter design,~\Eq{eq:uc} is first linearized as 
\begin{equation}\label{eq:uc_l}
    \pt\bm u_c\approx-\bm k_v\left(\psi_0\pt\omega_c+\frac{\omega_0}{\psi_0}\bm \psi_0^T\pt{\bm \psi}\right), 
\end{equation} 
where $\psi_0=V^*/\omega_0$, $\bm \psi_0=\bm \psi^*$. The approximation is resulted from $\pt\hat{\bm \psi}\approx\pt\bm \psi$, which is reasonable with high bandwidth of virtual flux estimation. PCC voltage magnitude~\Eq{eq:V} is linearized as 
\begin{equation}\label{eq:V_l}
    \pt V=\frac{1}{V^*}\bm u_c^{*T}\pt\bm u_c. 
\end{equation} 
Substitute~\Eq{eq:uc_l} into~\Eq{eq:V_l}, it is obvious that the closed-loop poles of $\pt V$ can be equivalently assigned through pole placement for $\pt\bm \psi$. To this end,~\Eq{eq:psi} is linearized as 
\begin{equation}\label{eq:ptpsidot}
    \pt\dot{\bm \psi}=-\omega_0\bm J\pt\bm \psi+\pt\bm u_c-\bm J\bm \psi_0\pt\omega_c. 
\end{equation} 
Substitute~\Eq{eq:uc_l} into~\Eq{eq:ptpsidot}, 
\begin{equation}
\begin{aligned} 
    \pt\dot{\bm \psi}=&-\left(\omega_0\bm J+\bm k_v\bm w\right)\pt\bm \psi-\left(\bm J\bm \psi_0+\psi_0\bm k_v\right)\pt\omega_c, 
    \end{aligned} 
\end{equation}
where $\bm w=\ma{0&-\omega_0}$. Therefore, $\bm k_v$ can be designed through pole placement such that 
\begin{equation}\label{eq:sigmav}
    \text{eig}(-\omega_0\bm J-\bm k_v\bm w)=\bm \sigma_v, 
\end{equation}
where $\bm \sigma_v$ specifies two poles on left half of $s$-plane.

\subsection{Derivation of closed-loop system} 
\begin{figure}[!tbp]
\includegraphics[width=0.5\textwidth,trim={3cm 4cm 3cm 3cm},clip]{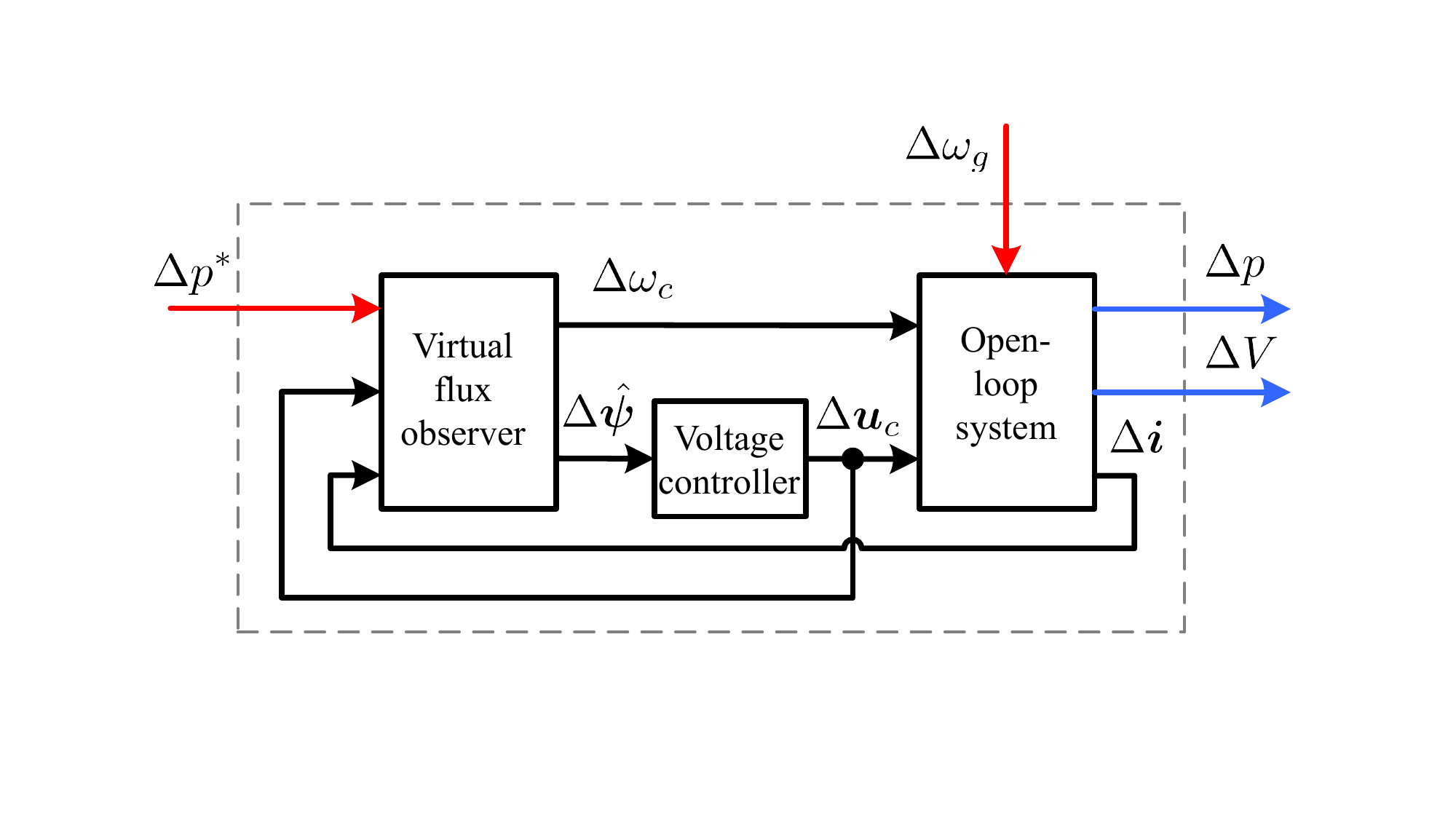}
\caption{Small-signal model of the closed-loop system. }\label{fig:fig4}
\end{figure}

The small-signal model of the closed-loop system is shown in Fig.~\ref{fig:fig4}. The small-signal state-space models of the component systems can be derived conveniently from~\Eq{eq:ob},~\Eq{eq:uc} and~\Eq{eq:ic}. Then, the subsystems are connected according to the input-output relations specified in Fig.~\ref{fig:fig4} to obtain the closed-loop model. 

Due to space limitations, the mathematical details are not presented in this article. For stability analysis, the eigenvalues of the closed-loop system matrix are used as indicators. Numerical results of closed-loop stability will be presented in the following section. 

\begin{figure}[!tbp] 
\centering 
\includegraphics[width=0.48\textwidth]{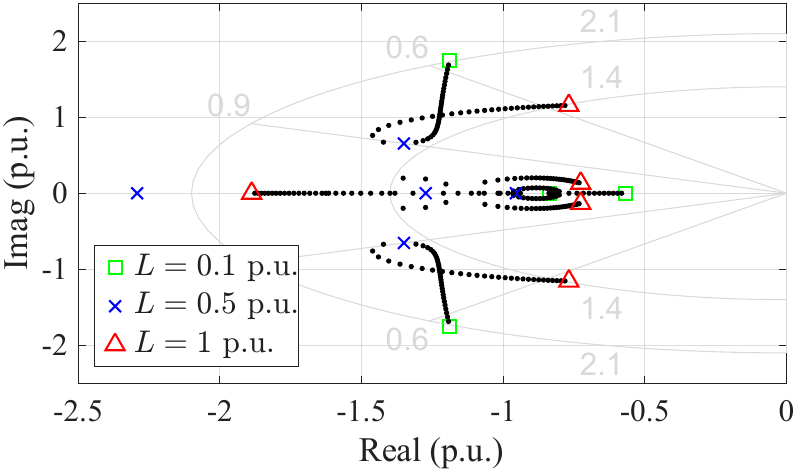}
\caption{Closed-loop pole trajectories of the proposed method under total impedance variation from \SI[parse-numbers = false]{L=0.1}{\pu} to \SI[parse-numbers = false]{L=1}{\pu}. }\label{fig:simF}
\end{figure} 

\section{Simulation results}
In simulations, stability robustness of the proposed method is demonstrated in both frequency domain and time domain. The rated parameters of GCC and the base values are given in Table~\ref{t:1}.

\subsection{Parameter tuning} 
The control parameters of the proposed method are designed as follows: 
\begin{enumerate}
    \item The closed-loop poles of flux estimation $\bm \sigma_o$ are assigned by $\bm k_o$ around $s=-2.5\omega_0$, through~\Eq{eq:sigmao}. 
    \item In~\Eq{eq:kp}, $\bm k_p$ is given by $\zeta=0.9$ and $\omega_s=1.5\omega_0$. 
    \item The closed-loop poles of voltage control $\bm \sigma_v$ are assigned by $\bm k_v$ around $s=-\omega_0$, through~\Eq{eq:sigmav}. 
\end{enumerate} 
The operating point $\delta^*$ used in control parameter calculations is given by~\Eq{eq:ptodelta} where \SI[parse-numbers = false]{p^*=1}{\pu} and \SI[parse-numbers = false]{L_0=0.5}{\pu}. The same control parameters are used throughout the simulation and experimental studies of this paper. 

The varying grid strength is configured by varying total impedance $L$ from \SI{0.1}{\pu} to \SI{1}{\pu}, \ie short-circuit ratio (SCR) from $10$ to $1$. 

\subsection{Result and discussion} 

Fig.~\ref{fig:simF} shows the pole trajectories of the small-signal closed-loop system when \SI[parse-numbers = false]{L_0=0.5}{\pu} is used in the proposed controller, while the actual impedance $L$ varies from $0.1$ to $1$ p.u.. It can be observed that the poles experience small range excursions on the left half plane, thus maintaining stability. When the grid strength is very strong or very weak, the dynamic response of the proposed method will be slower than designed, as the dominant pole moves from $s=-\omega_0$ when \SI[parse-numbers = false]{L=0.5}{\pu} to $s=-0.6\omega_0$ when \SI[parse-numbers = false]{L=1}{\pu} and \SI{0.1}{\pu}. 

\begin{table}[!t]\caption{Rated parameters of the experimental platform}\label{t:1}
\centering
\begin{tabular}{lcc}
\hline
Parameter                    & Actual value & Per unit value \\ \hline
Rated power                  & \SI{20}{\kilo\voltampere}           & \SI{1}{\pu}              \\
Rated voltage                & \SI[parse-numbers = false]{\sqrt{2/3}\times 380}{\volt}          & \SI{1}{\pu}              \\
Rated current                & \SI[parse-numbers = false]{\sqrt{2}\times 30}{\ampere}           & \SI{1}{\pu}              \\
Base inductance               & \SI{23}{\milli\henry}          & \SI{1}{\pu}              \\
Base frequency        & \SI{50}{\hertz}           & \SI{1}{\pu}              \\
Filter impedance               & \SI{2.5}{\milli\henry}          & \SI{0.1}{\pu}              \\
Switching/Sampling frequency & \SI{10}{\kilo\hertz}          & \SI{200}{\pu}            \\
Rated DC-link voltage & \SI[parse-numbers = false]{750}{\volt}          & \SI{2.4}{\pu}              \\
\hline
\end{tabular}
\end{table} 

\section{Experimental results} 
\subsection{Experimental setup}

\begin{figure}[!tbp] 
\centering 
\includegraphics[width=0.5\textwidth,trim={5cm 3.5cm 5cm 3cm},clip]{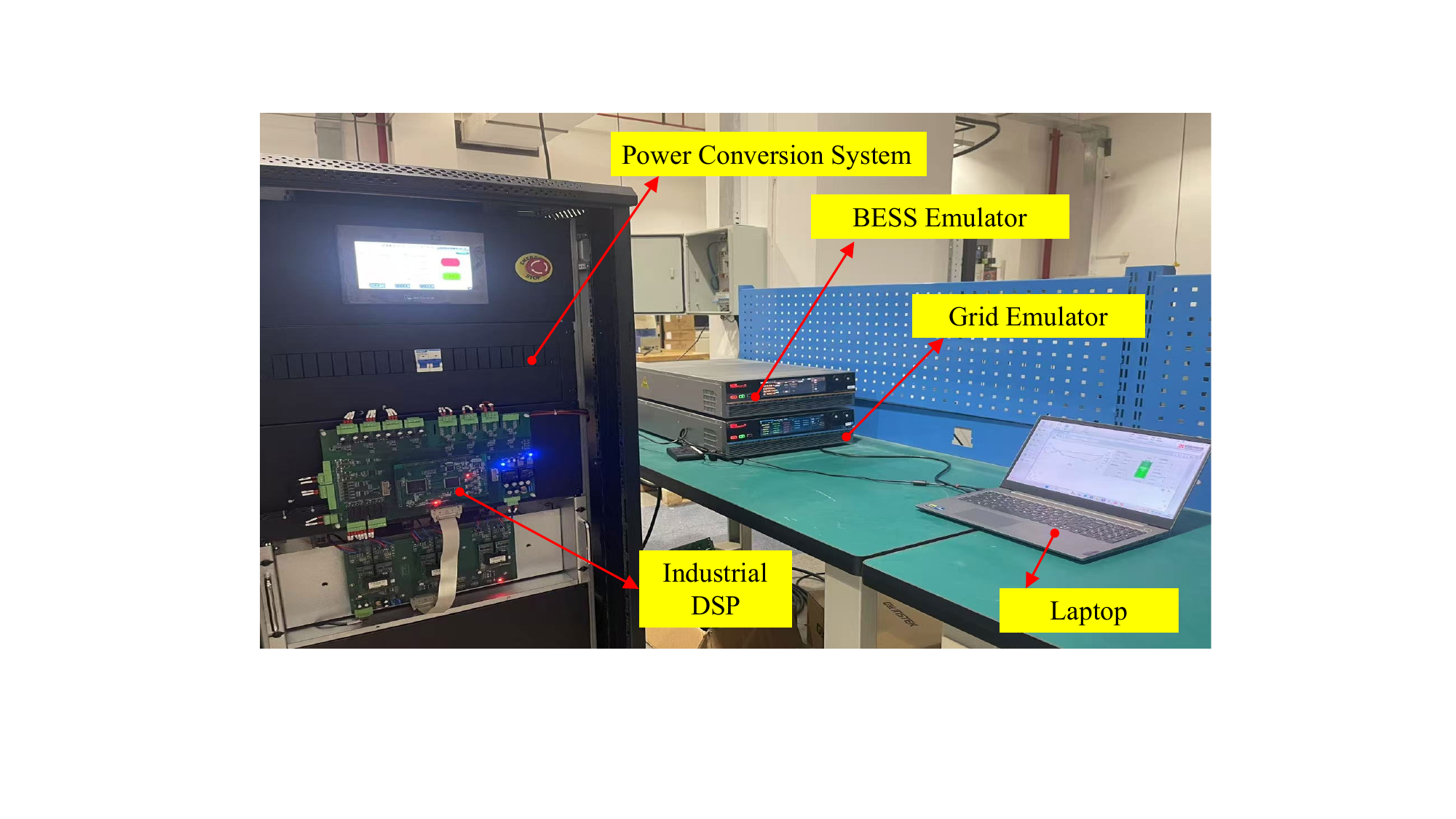}
\caption{Experimental platform with the \SI{20}{\kilo\voltampere} PCS, grid emulator and BESS emulator. }\label{fig:platform}
\end{figure} 
In the following, the proposed GFM controller is validated on a \SI{20}{\kilo\voltampere} power conversion system (PCS) experimental platform, as shown in Fig.~\ref{fig:platform}. The PCS mainly comprises a 3-phase 2-level power converter and a digital signal processor (DSP) TMS320F28335, on which the proposed GFM controller is implemented. The DC-side of the PCS is connected to a battery energy storage system (BESS) emulator. The AC-side is connected to a grid emulator. A variable impedance cabinet serves as variable grid impedance. The parameters of the experimental platform are given in Table~\ref{t:1}. The results are measured with an oscilloscope at sampling rate \SI{25}{\kilo\hertz}.  

For comparison, the reference feedforward power synchronization control (RFPSC) method~\cite{RFPSC2020} is also implemented in the experiments. The control parameters are selected as \SI[parse-numbers = false]{R_a=0.2}{\pu} and \SI[parse-numbers = false]{\omega_b=0.1}{\pu}~\cite{Robust2019,RFPSC2020}.

\subsection{Results and discussions} 

\begin{figure}[!tbp] 
\centering 
\includegraphics[width=0.5\textwidth]{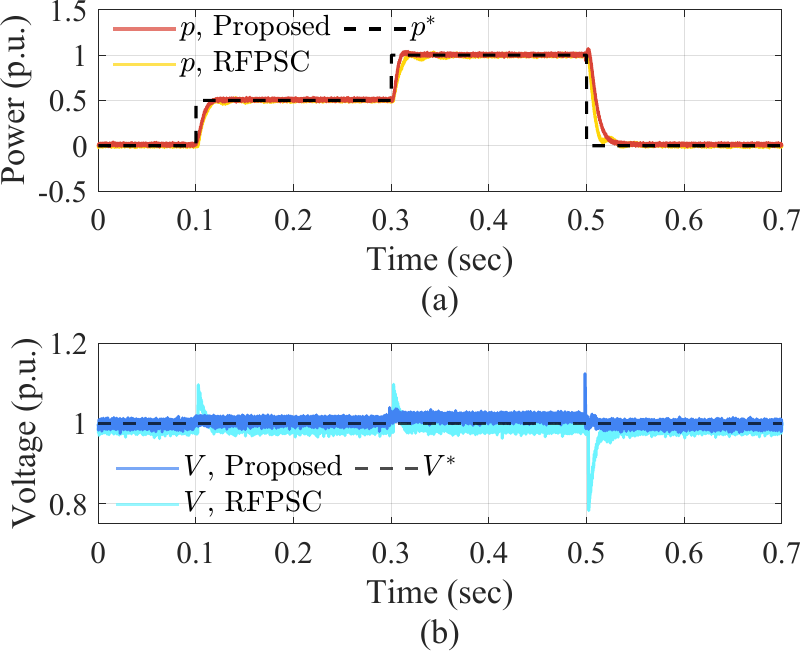}
\caption{Comparison of active power reference tracking performance between the proposed method and RFPSC when \SI[parse-numbers = false]{L=0.5}{\pu}. }\label{fig:exp1}
\end{figure} 
First, the control performance of the proposed method is compared with RFPSC when \SI[parse-numbers = false]{L=L_0=0.5}{\pu} in Fig.~\ref{fig:exp1}. The active power set-point is first stepped up to \SI{0.5}{\pu}, then to \SI{1}{\pu}, and finally stepped down to 0. From Fig.~\ref{fig:exp1}(a), both methods track the power reference accurately. Their dynamic responses are also similar, with no overshoot and the settling time is around \SI{0.02}{\second}. However, the proposed method is more advantageous than RFPSC in PCC voltage response, as shown in Fig.~\ref{fig:exp1}(b). With the proposed method, there are no obvious voltage peaks during active power step-ups. In contrast, RFPSC produces \SI{0.1}{\pu} increases in PCC voltage at \SI{0.1}{\second} and \SI{0.3}{\second}, which only settle down after \SI{0.02}{\second} along with the transients of active power. Similarly, during active power step-down, PCC voltage drops lower than \SI{0.8}{\pu} with RFPSC, and takes \SI{0.05}{\second} to recover. With the proposed method, it is increased to \SI{1.1}{\pu} and restored within \SI{0.01}{\second}. 

\begin{figure}[!tbp] 
\centering 
\includegraphics[width=0.5\textwidth]{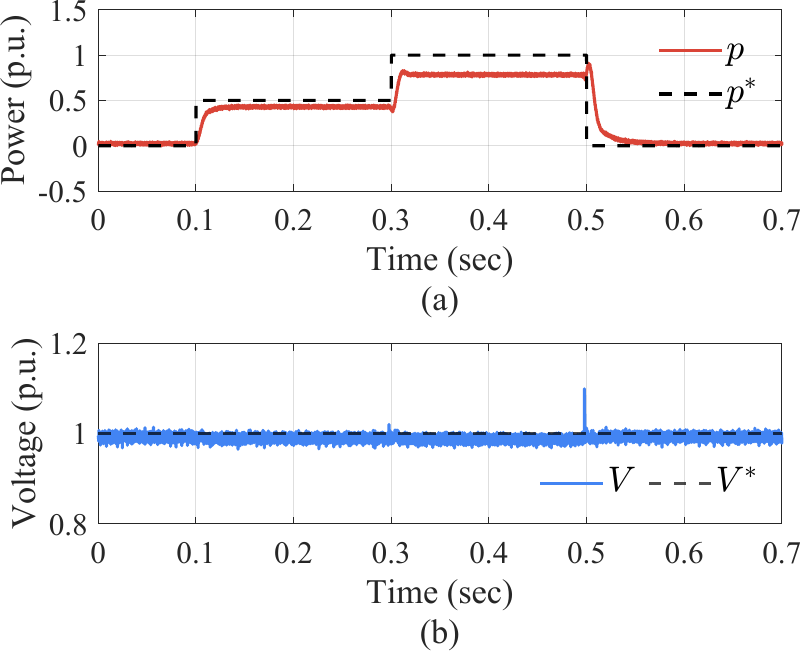}
\caption{Control performance of the proposed method when \SI[parse-numbers = false]{L=1}{\pu}. }\label{fig:exp2}
\end{figure} 

Next, stability robustness of the proposed method to grid strength variation is validated. Fig.~\ref{fig:exp2} shows the control performance of the proposed method under weak grid condition \SI[parse-numbers = false]{L=1}{\pu}. In Fig.~\ref{fig:exp2}(a), active power offsets occur due to plant parameter mismatch, which is common for observer-based controller and can be eliminated by identification of actual parameters or disturbance compensation algorithms. Meanwhile, the dynamic performance is robust to total impedance variation, by comparing the active power transients in Fig.~\ref{fig:exp2}(a) and Fig.~\ref{fig:exp1}(a). As $L$ changes from design parameter \SI{0.5}{\pu} to \SI{1}{\pu}, the active power response remains damped with slightly increased settling time. In Fig.~\ref{fig:exp2}(b), the advantage of stiff PCC voltage magnitude of the proposed method is also preserved under weak grid condition.

\begin{figure}[!tbp] 
\centering 
\includegraphics[width=0.5\textwidth]{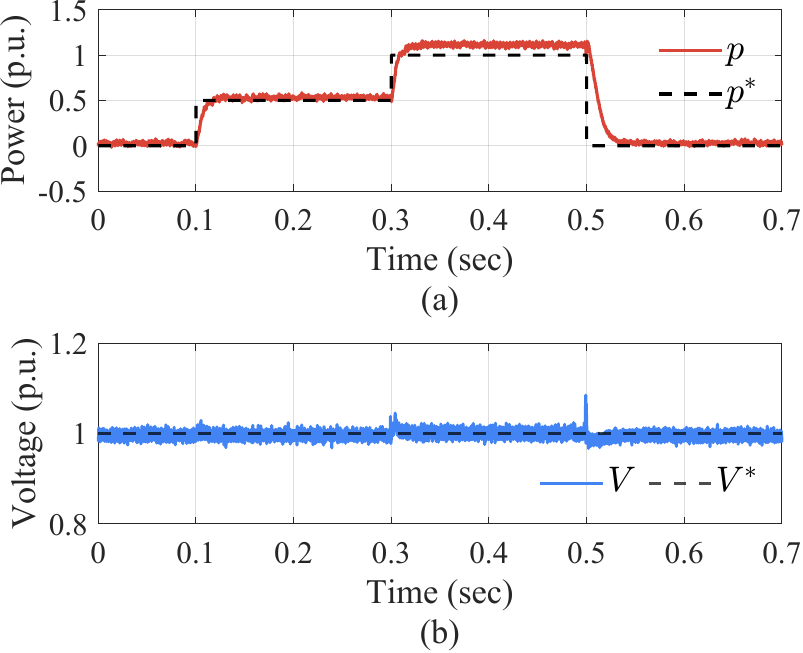}
\caption{Control performance of the proposed method when \SI[parse-numbers = false]{L=0.1}{\pu}. }\label{fig:exp3}
\end{figure}

\begin{figure}[!tbp] 
\centering 
\includegraphics[width=0.5\textwidth]{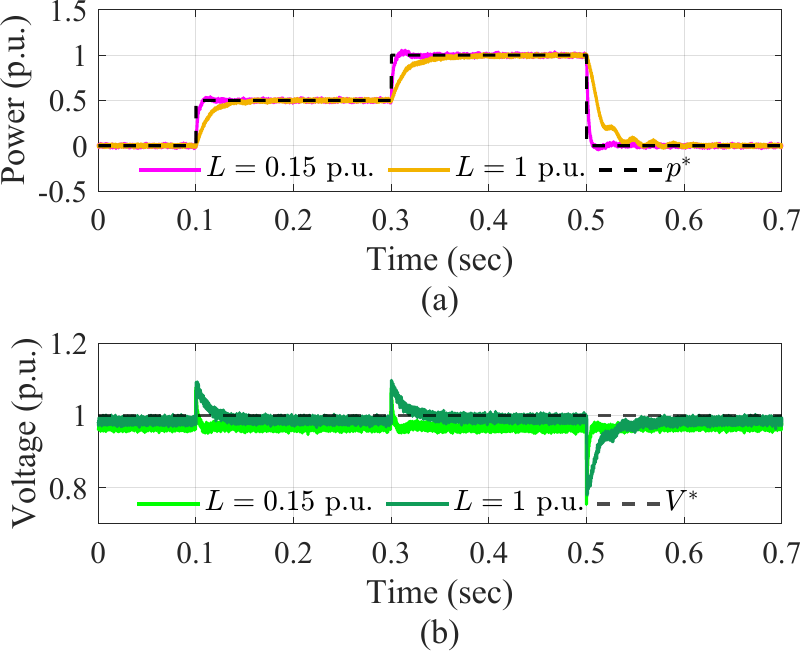}
\caption{Control performance of RFPSC when \SI[parse-numbers = false]{L=0.15}{\pu} and \SI[parse-numbers = false]{1}{\pu}. }\label{fig:exp4}
\end{figure}

Similar conclusions can be drawn from Fig.~\ref{fig:exp3}, where total impedance changes to \SI[parse-numbers = false]{L=0.1}{\pu} to form weak grid condition. Despite the \SI{0.1}{\pu} offset in active power, the transient response in Fig.~\ref{fig:exp3}(a) is almost identical with the accurate impedance case in Fig.~\ref{fig:exp1}(a). In Fig.~\ref{fig:exp3}(b), PCC voltage magnitude exhibits short transients of \SI{0.01}{\second} during step-up and step-down of \SI{1}{\pu} active power, which is slightly increased compared to medium and weak grid strengths.

To better demonstrate the robustness of the proposed method, the control performances of RFPSC under weak and strong grid conditions are compared in Fig.~\ref{fig:exp4} for reference. The deviation of dynamic response of the two cases is evident. The settling time of active power and PCC voltage is increased from \SI{0.01}{\second} in strong grid condition, to \SI{0.05}{\second} in weak grid condition. In contrast, with the proposed method, the settling time of active power remains around \SI{0.02}{\second} across the range, and the stiff voltage source characteristic of PCC voltage is maintained in weak and strong grid alike.

\begin{figure}[!tbp] 
\centering 
\includegraphics[width=0.5\textwidth]{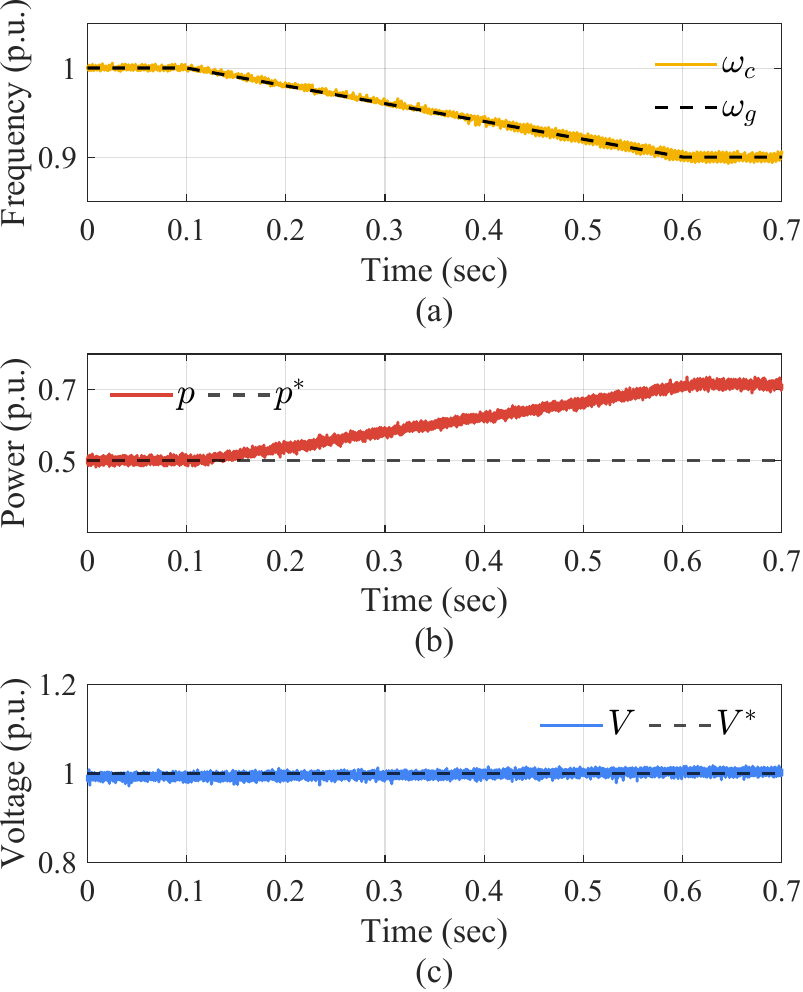}
\caption{Synchronization and control performance of the proposed method when grid frequency is ramped from \SI{50}{\hertz} to \SI{45}{\hertz}, \SI[parse-numbers = false]{L=0.5}{\pu}. }\label{fig:exp5}
\end{figure} 

Finally, synchronization and output response of the proposed method during grid frequency variation is demonstrated in Fig.~\ref{fig:exp5}. Fig.~\ref{fig:exp5}(a) shows that the rotating frequency of controller coordinate frame $\omega_c$ closely follows the decreasing grid frequency $\omega_g$ with no visible transient. In response to the frequency drop, active power output of GCC is ramped up from its set-point of \SI{0.5}{\pu} to \SI{0.7}{\pu} smoothly, as shown in Fig.~\ref{fig:exp5}(b), while the PCC voltage is only slightly increased from the set-point in Fig.~\ref{fig:exp5}(c). Notably, the active power offset induced by frequency mismatch corresponds to a droop coefficient of \SI[parse-numbers = false]{D_p=2}{\pu} in the proposed GFM controller. 

In conclusion, the experimental results demonstrate the robust dynamic performance of the proposed method across a wide range of grid strengths. With the proposed GFM controller, the GCC exhibits stiff voltage source behavior and droop characteristics. 
\bibliographystyle{Bibliography/IEEEtranTIE}
\bibliography{Bibliography/IEEEabrv,Refs}

@ARTICLE{ORathnayake2021,
  author={Rathnayake, Dayan B. and Akrami, Milad and Phurailatpam, Chitaranjan and Me, Si Phu and Hadavi, Sajjad and Jayasinghe, Gamini and Zabihi, Sasan and Bahrani, Behrooz},
  journal={IEEE Access}, 
  title={Grid Forming Inverter Modeling, Control, and Applications}, 
  year={2021},
  volume={9},
  number={},
  pages={114781-114807}
}

@ARTICLE{ORosso2021,
  author={Rosso, Roberto and Wang, Xiongfei and Liserre, Marco and Lu, Xiaonan and Engelken, Soenke},
  journal={{IEEE} Open J. Ind. Appl.}, 
  title={Grid-Forming Converters: Control Approaches, Grid-Synchronization, and Future Trends—A Review}, 
  year={2021},
  volume={2},
  number={},
  pages={93-109}
}

@article{OLasseter2020,
   author = {Lasseter, Robert H. and Chen, Zhe and Pattabiraman, Dinesh},
   title = {Grid-Forming Inverters: A Critical Asset for the Power Grid},
   journal = {{IEEE} J. Emerg. Sel. Topics Power
Electron. },
   volume = {8},
   number = {2},
   pages = {925-935},
   year = {2020},
   type = {Journal Article}
}

@ARTICLE{Li2022,
  author={Li, Yitong and Gu, Yunjie and Green, Timothy C.},
  journal={{IEEE} Trans. Power Syst.}, 
  title={Revisiting Grid-Forming and Grid-Following Inverters: A Duality Theory}, 
  year={2022},
  volume={37},
  number={6},
  pages={4541-4554}
}

@ARTICLE{Wang2020,
  author={Wang, Xiongfei and Taul, Mads Graungaard and Wu, Heng and Liao, Yicheng and Blaabjerg, Frede and Harnefors, Lennart},
  journal={{IEEE} Open J. Ind. Appl.}, 
  title={Grid-Synchronization Stability of Converter-Based Resources—An Overview}, 
  year={2020},
  volume={1},
  number={},
  pages={115-134}
}

@ARTICLE{Gao2024,
  author={Gao, Xian and Zhou, Dao and Anvari-Moghaddam, Amjad and Blaabjerg, Frede},
  journal={{IEEE} Trans. Ind. Appl.}, 
  title={Stability Analysis of Grid-Following and Grid-Forming Converters Based on State-Space Modelling}, 
  year={2024},
  volume={60},
  number={3},
  pages={4910-4920}
}

@article{Zhang2010,
   author = {Zhang, Lidong and Harnefors, Lennart and Nee, Hans-Peter},
   title = {Power-Synchronization Control of Grid-Connected Voltage-Source Converters},
   journal = {{IEEE} Trans. Power Syst.},
   volume = {25},
   number = {2},
   pages = {809-820},
   year = {2010},
   type = {Journal Article}
}

@article{Robust2019,
   author = {Harnefors, Lennart and Hinkkanen, Marko and Riaz, Usama and Rahman, F. M. Mahafugur and Zhang, Lidong},
   title = {Robust Analytic Design of Power-Synchronization Control},
   journal = {{IEEE} Trans. Ind. Electron.},
   volume = {66},
   number = {8},
   pages = {5810-5819},
   year = {2019},
   type = {Journal Article}
}

@article{Universal2021,
   author = {Harnefors, Lennart and Kukkola, Jarno and Routimo, Mikko and Hinkkanen, Marko and Wang, Xiongfei},
   title = {A Universal Controller for Grid-Connected Voltage-Source Converters},
   journal = {{IEEE}
Trans. Emerg. Sel. Topics Power Electron.},
   volume = {9},
   number = {5},
   pages = {5761-5770},
   year = {2021},
   type = {Journal Article}
}

@article{RFPSC2020,
   author = {Harnefors, Lennart and Rahman, F. M. Mahafugur and Hinkkanen, Marko and Routimo, Mikko},
   title = {Reference-Feedforward Power-Synchronization Control},
   journal = {{IEEE} Trans. Power Electron.},
   volume = {35},
   number = {9},
   pages = {8878-8881},
   year = {2020},
   type = {Journal Article}
}

@article{LiuUVC2024,
   author = {Liu, T. and Wang, X.},
   title = {Unified Voltage Control for Grid-Forming Inverters},
   journal = {{IEEE} Trans. Ind. Electron.},
   volume = {71},
   number = {3},
   pages = {2578-2589},
   year = {2024},
   type = {Journal Article}
}

@article{Zhao2024,
   author = {Zhao, Fangzhou and Zhu, Tianhua and Harnefors, Lennart and Fan, Bo and Wu, Heng and Zhou, Zichao and Sun, Yin and Wang, Xiongfei},
   title = {Closed-Form Solutions for Grid-Forming Converters: A Design-Oriented Study},
   journal = { {IEEE} Open J. Power Electron.,},
   volume = {5},
   pages = {186-200},
   year = {2024},
   type = {Journal Article}
}

@article{Zhao2024l,
   author = {Zhao, Fangzhou and Zhu, Tianhua and Li, Zejie and Wang, Xiongfei},
   title = {Low-Frequency Resonances in Grid-Forming Converters: Causes and Damping Control},
   journal = {{IEEE} Trans. Power Electron.},
   volume = {39},
   number = {11},
   pages = {14430-14447},
   year = {2024},
   type = {Journal Article}
}

@article{Nurminen2024,
   author = {Nurminen, T. and Mourouvin, R. and Hinkkanen, M. and Kukkola, J.},
   title = {Multifunctional Grid-Forming Converter Control Based on a Disturbance Observer},
   journal = {{IEEE} Trans. Power Electron.},
   volume = {39},
   number = {10},
   pages = {13023-13032},
   year = {2024},
   type = {Journal Article}
}

@article{Nurminen2025,
   author = {Nurminen, T. and Mourouvin, R. and Hinkkanen, M. and Kukkola, J.},
   title = {Design and Sensitivity Analysis of Grid-Forming Converter Control Based on a Disturbance Observer},
   journal = {{IEEE} Trans. Ind. Appl.},
   pages = {1-13},
   year = {2025},
   type = {Journal Article}
}

@INPROCEEDINGS{Nurminen2023,
  author={Nurminen, Tuure and Mourouvin, Rayane and Hinkkanen, Marko and Kukkola, Jarno and Routimo, Mikko and Vilhunen, Antti and Harnefors, Lennart},
  booktitle={2023 IEEE Belgrade PowerTech}, 
  title={Observer-Based Power-Synchronization Control for Grid-Forming Converters}, 
  year={2023},
  volume={},
  number={},
  pages={1-6}
}

@ARTICLE{Marko2018,
  author={Hinkkanen, Marko and Saarakkala, Seppo E. and Awan, Hafiz Asad Ali and Mölsä, Eemeli and Tuovinen, Toni},
  journal={{IEEE} Trans. Ind. Appl.}, 
  title={Observers for Sensorless Synchronous Motor Drives: Framework for Design and Analysis}, 
  year={2018},
  volume={54},
  number={6},
  pages={6090-6100}
}

@ARTICLE{Flux2022,
  author={Zhang, Mingming and Xia, Binyu and Zhang, Jun},
  journal={{IEEE} Trans. Energy Convers.}, 
  title={Parameter Design and Convergence Analysis of Flux Observer for Sensorless {PMSM} Drives}, 
  year={2022},
  volume={37},
  number={4},
  pages={2512-2524}
}

\end{document}